\documentstyle{article}
\begin{document}

\begin{center}
{\Large Hydrodynamic accretion onto rapidly rotating Kerr black
hole}
\end{center}

\vspace{1cm}
\begin{center}
{\large V.I.Pariev}
\end{center}

\begin{center}
{\it Lebedev Physical Institute, Leninsky Prospect 53, Moscow
117924, Russia}
\end{center}

\vspace{5cm}
\begin{center}
{\Large Abstract}
\end{center}
Bondi type hydrodynamic accretion of the surrounding matter onto Kerr
black hole with an arbitrary rotational parameter is considered. The
effects of viscosity, thermal conductivity and interaction with radiation
field are neglected. The black hole is supposed to be at rest with respect
to matter at infinity. The flow is adiabatic and has no angular momentum.
The fact that usually in astrophysics substance far from the black hole
has nonrelativistic temperature introduces small parameter to the problem
and allows to search for the solution as a perturbation to the
accretion of a cold, that is dust--like, matter. However, far from the
black hole on the scales of order of the radius of the sonic surface the
expansion must be performed with respect to Bondi spherically symmetrical
solution for the accretion on a Newtonian gravitating centre. The
equations thus obtained are solved analytically. The conditions of the
regularity of the solution at the sonic surface and at infinity allow to
specify unique solution, to find the shape of the sonic surface and to
determine the corrections to Bondi accretion rate.

\vspace{1cm}


\vspace{10mm}
\begin{center}
Submitted to {\it Monthly Notices}
\end{center}

\newpage

\section{Introduction}

The problem of the matter accretion onto gravitating objects is the
classical problem of astrophysics~(Zeldovich \& Novikov~1967; Shapiro \&
Teukolsky~1983). The accretion is widely engaged for the explanation of
various kinds of astrophysical phenomena. It is believed that this process
is basic to understanding the nature of active galactic nuclei,
quasars, galactic X-ray sources, the formation of extragalactic jets and
jets from some galactic objects of stellar masses~(i.e., Begelman,
Blandford \& Rees~1984; Camenzind~1990; Lipunov~1992). Usually, when
considering accretion with astrophysical implications, the authors
either worked in the frame of a disk accretion model or considered
spherically symmetrical case, which enabled them without excessive details
to estimate the accretion rate and the emerging radiation due to the
heating of infalling matter (Shapiro \& Teukolsky~1983). Bondi~(1952) was
the first who considered hydrodynamic isoentropic stationary flow for
polytropic equation of state and showed that, being subsonic at infinity,
flow must cross the sonic surface and become supersonic near the black
hole, the accretion rate being determined by the requirement of smoothness
of the solution at the sonic surface.

Present paper investigates the adiabatic stationary accretion onto
a rotating black hole. Let us first note that the luminosity of the
infalling matter itself is much less than the Eddington limit in
quasispherical case~(Shapiro \& Teukolsky~1983), so,  despite that our
approach does not involve interaction with photons, it can be applied for
the accretion by an isolated black hole embedded in interstellar gas or
molecular cloud. Probably Shapiro~(1974) was the first who considered
accretion onto a Kerr black hole. He noticed that in the case of zero
temperature the situation is similar to the accretion of dust, so  the
particles would follow geodesic lines in Kerr space-time and flow lines
would simply coincide with them. As it can be verified directly from
the equations for the test particle motion in Kerr geometry (e.g.,
Novikov~\& Frolov, 1986), a particle, resting with respect to the black
hole at infinity, will move purely radially (in Boyer--Lindquist
coordinates) and experience dragging in $\phi$ direction with
Lense--Thirring velocity, while any additional impact to it in radial
direction causes it to deflect in $\theta$ direction. Therefore, we
already have the accretion pattern for cold matter. One should also expect
that the slight heating of the gas far from the black hole, such that the
gas molecules thermal velocity remains to be much smaller than the
velocity of light, would not alter the flow picture drastically. Having
had noticed all that, Shapiro came to calculation of the emerging
bremsshtralung radiation from almost cold infalling material.  Further
step was undertaken by Petrich, Shapiro \& Teukolsky in 1988. They found
analytical solution for arbitrary fast rotating Kerr black hole moving
through the surrounding gas along its rotational axis, but for equation of
state $p=\varrho$ with the sound velocity $c_s$ equal to the light
velocity everywhere, which looks very artificial.

Recently Beskin \& Pidoprigora~(1995) (further referred to as BP)
considered the accretion for polytropic equation of state by slowly
rotating black hole.  They used small ratio of the rotational
parameter~$a$ to the black hole mass $\cal M$ as a small parameter to
linearize hydrodynamic equations and to search for the solution as a small
perturbation to the spherically symmetrical accretion by Schwarzschild
black hole. The aim of the present work is to find the solution for the
Kerr black hole with an {\it arbitrary value of~$a$}. As in BP the black
hole is supposed to be at rest with respect to the gas at infinity, flow
is adiabatic and without angular momentum with respect to the black hole
at infinity, matter has uniform temperature and density far from the black
hole. Under these conditions flow must be axisymmetrical with the symmetry
axis being the rotational axis of the black hole.

For solving the problem formulated we use the equation of flow lines
equilibrium (stream equation), which is nonlinear differential equation of
mixed type in flux function~$\Phi$ and is of Grad--Shafranov type.
Its general form for Kerr metric has been derived in Beskin \&
Pariev~(1993). In section~2 basic quantities describing the flow are
introduced and the equilibrium equation is formulated. As it is pointed
above this equation must have solution corresponding to the accretion of
the cold matter having zero pressure~($p=0$).  It occurred to be really
true as have been verified in BP.  The fact that
usually in astrophysics substance far from the black hole has
nonrelativistic temperature introduces small parameter to the
problem~---~the ratio of the temperature at infinity to the rest mass of a
particle, and allows us to search for the solution as a perturbation to
 the accretion of a cold matter. Stream equation can be linearized with
respect to this parameter and the linearized equation occurred to be
solvable analytically. This has been done in section~3. However,
the solution for cold matter cannot be considered as a zero order solution
for the expansion on the distances from the black hole comparable to the
radius of the sonic surface (which is much greater than the radius of the
event horizon), because the appearance of the sonic surface itself is
essentially connected with the nonzero temperature of the matter. In
section~4 we have constructed solution of the expanded stream equation in
that region using as a zero order approximation the solution for Bondi
accretion flow onto Newtonian gravitating centre of mass~$\cal M$.
Matching of the two solutions allows to determine the accretion pattern
uniquely.

Construction of the analytical shock--free solution for the hydrodynamic
accretion onto Kerr black hole leads us to the generalization for
rapid rotation of the result of the work BP that neither the rotation
of a black hole nor its slow motion change drastically the nature of the
accretion from the spherically symmetrical smooth flow.

\section{Basic equations}

Let us consider an axisymmetrical steady--state flow in the Kerr
space--time with the metric in Boyer--Lindquist coordinates $x^0=t$,
$x^1=r$, $x^2=\theta$, $x^3=\phi$:  \begin{equation} ds^2=-\alpha^2
dt^2+g_{ik}(dx^i+\beta^i dt)(dx^k+\beta^k dt)\mbox{,} \label{asc1}
\end{equation}
where
\begin{equation}
\alpha=\frac{\rho}{\Sigma}\sqrt{\Delta}\label{asc2}
\end{equation}
is the lapse function ($\alpha=\Delta=0$ at the event horizon),
\begin{equation}
\beta^r=\beta^{\theta}=0, \qquad
\beta^{\phi}=-\omega=-\frac{2a{\cal M}r}{\Sigma^2}\mbox{,}\label{asc3}
\end{equation}
$\omega$ is the angular velocity of Lense--Thirring precession, for all
$i\neq k$ $g_{ik}=0$,
\begin{equation}
g_{rr}=\rho^2/\Delta, \qquad g_{\theta\theta}=\rho^2,
\qquad g_{\phi\phi}
=\varpi^2\mbox{,}\label{asc4}
\end{equation}
and
\begin{eqnarray}
&& \Delta=r^2+a^2-2{\cal M}r, \qquad \rho^2=r^2+a^2\cos^2 \theta, \nonumber\\
&& \Sigma^2=(r^2+a^2)^2-a^2\Delta\sin^2\theta, \qquad \varpi=\frac{\Sigma}
{\rho}\sin\theta\mbox{.}\label{asc5}
\end{eqnarray}
Here ${\cal M}$ is the mass of a black hole, $a=J/{\cal M}$ is the
specific angular momentum of a black hole. Throughout all the paper we
will adopt the notations used in  Beskin~\& Pariev~(1993) and in BP. We
will use units where $c=1$ and $G=1$, that is all physical quantities have
dimensions of some power of the unit of length. We follow the '3+1'--
splitting of the space--time described in the book by Thorne et
al.~(1986).  All physical quantities considered here are measured by Zero
Angular Momentum Observers (ZAMO), rotating around the black hole at
a constant radius $r$ with Lense--Thirring angular velocity
$d\phi/dt=\omega$.  All 3-dimensional vectors and tensors are referred to
3-dimensional 'absolute' space with
metric $g_{ik}$~(\ref{asc4}) orthogonal to the world lines of ZAMOs.
It is convenient to use the following
orthonormal triad in this 3-space
$$ {\bf
e}_{\hat\phi}=\frac{\sqrt{\Delta}}{\rho}\frac{\partial}{\partial
r}\mbox{,}\qquad {\bf e}_{\hat\theta}=\frac{1}{\rho}\frac{\partial}{
\partial\theta}\mbox{,} \qquad {\bf e}_{\hat\phi}=\frac{1}{\varpi}
\frac{\partial}{\partial \phi}\mbox{.} $$

Axial symmetry and time independence allow us to introduce flux function
$\Phi(r,\theta)$ defined as
\begin{equation}
\alpha n {\bf u}_P=\frac{1}{2\pi\varpi}(\nabla \Phi\times{\bf e}_
{\hat\phi})\mbox{.}\label{asc6}
\end{equation}
Here $n$ is the total particle density in the comoving reference frame,
${\bf u}_P$ is the poloidal (that is in the directions of ${\bf
e}_{\hat\theta}$ and ${\bf e}_{\hat r}$) component of the four velocity
$u^i$. Due to the definition of $\Phi(r,\theta)$ by formula~(\ref{asc6})
the continuity equation {\boldmath$\nabla$}$\cdot(\alpha n{\bf u})=0$ is
satisfied automatically. Because of $({\bf u}\mbox{\boldmath$\nabla$}
\Phi)=0$ the flux function $\Phi(r,\theta)$ is constant along each stream
line, and the quantity $\displaystyle \int\limits_S\,
d\Phi=\int\limits_S \alpha n\,{\bf u}d{\bf S}$ is the total flux
of the particles through an area~$S$. To avoid singularities on the
rotational axis~$\theta=0$ flux function must satisfy two conditions
$\displaystyle \left.\Phi\right|_{\theta=0}=0$ and $\displaystyle
\left.\frac{\partial \Phi}{\partial\theta}\right|_{\theta=0}=
\left.\frac{\partial \Phi}{\partial\theta}\right|_{\theta=\pi}=0 $. Note
also that the value $\displaystyle -\left.\Phi\right|_{\theta=\pi}$ is the
total accretion rate (with positive sign). Particularly, in the case of
spherical symmetry $\Phi=\Phi_0(1-\cos\theta)$, and the accretion rate is
$-2\Phi_0$.

Following the tradition we adopt polytropic equation of state
\begin{equation}
p=k(s)n^\Gamma\mbox{,}\label{asc7}
\end{equation}
where $p$ is the gas pressure, $1<\Gamma\leq 5/3$ is the polytropic index,
and $k(s)$ is the quantity depending only on the entropy per particle~$s$.
The flow is adiabatic, therefore,~$s$ is constant along stream lines
$\Phi(r,\theta)=\mbox{const}$, so $s=s(\Phi)$. Then, entalpy per particle
$\mu=(p+\varrho_m)/n$ (here $\varrho_m$ is the specific internal energy of
the gas) is
\begin{equation}
\mu=m+\frac{\Gamma}{\Gamma-1}k(s)n^{\Gamma-1}\mbox{,}\label{asc8}
\end{equation}
where $m$ is the mean rest mass of a particle. Accordingly, the sound
velocity is
\begin{equation}
c_s^2=\frac{1}{\mu}\left(\frac{\partial p}{\partial n}\right)_s=
\frac{\Gamma k(s)n^{\Gamma-1}}{\mu}\mbox{.}\label{asc9}
\end{equation}
All thermodynamic quantities can be expressed via the functions in~$s$
and~$n$ only. Moreover, in our problem gas temperature and density are
uniform on large distances from the black hole, so the entropy~$s$ is
constant along all stream lines, that is constant everywhere
$s=\mbox{const}$. Therefore, all thermodynamic quantities effectively
depend on only one variable and can be expressed via each other given the
value of~$s$.

The equations of motion are provided by four components of the
energy--momentum conservation low
\begin{equation}
T^{\hat\mu \hat\nu}_{\phantom{\hat\mu}\phantom{\hat\nu};\hat\nu}=0
\mbox{.}\label{asc10}
\end{equation}
The 0-component of~(\ref{asc10}) leads to the energy conservation along
each stream line (Bernoulli integral), which is manifested by that the
following quantity~$E$ depends only on~$\Phi$
\begin{equation}
E(\Phi)=\mu(\alpha\gamma+\varpi\omega u_{\hat\phi})\mbox{.}\label{asc11}
\end{equation}
Here $\gamma^2=1+u_{\hat P}^2+u_{\hat\phi}^2$ is the Lorentz factor of the
flow. The $\hat\phi$-component of~(\ref{asc10}) leads to the
conservation of the $z$-component of the angular momentum  along
each stream line
\begin{equation} L(\Phi)=\mu\varpi
u_{\hat\phi}\mbox{.}\label{asc12}
\end{equation}
For the problem in hand $L(\Phi)=0$, which means $u_{\hat\phi}=0$. Hence,
the gas corotates with ZAMOs with the angular velocity  just equal
to~$\omega$ as seen by an infinitly distant observer. At infinity
$\alpha=1$, $\gamma=\gamma_{\infty}=1$, and
$\mu=\mu_{\infty}=\mbox{const}$, therefore, one can see from~(\ref{asc11})
that $E(\Phi)=E=\mbox{const}$ along all stream lines and, hence,
everywhere. Now we can express all thermodynamic quantities and the
velocity~${\bf u}$ by means of $\Phi(r,\theta)$ and constants~$E$ and~$s$
only.  Using definition~(\ref{asc6}) and obvious relation
$\gamma^2=1+u_P^2$ and bearing in mind that $L=0$ one can obtain the
following equation
\begin{equation}
\varpi^2 E^2=\mu^2\left(\Delta\sin^2\theta+
\frac{|\mbox{\boldmath$\nabla$}\Phi|^2}{4\pi^2 n^2}\right)\label{asc13}
\mbox{,}
\end{equation}
which, being combined with the formula~(\ref{asc8}) for~$\mu$, provides
an implicit expression for~$n$ in terms of
$|\mbox{\boldmath$\nabla$}\Phi|^2$, $E$, and~$s$. Then,
all other characteristics of the flow can be obtained from~(\ref{asc6})
and~(\ref{asc8}).

The remaining two space poloidal components of the equation~(\ref{asc10})
should be decomposed into parts along the stream line, i.e.
perpendicular to $\mbox{\boldmath$\nabla$}\Phi$, and perpendicular to the
stream line, i.e. parallel to $\mbox{\boldmath$\nabla$}\Phi$. Component
along the stream line leads to the adiabatic condition $s=s(\Phi)$ and
gives no new information, because even the form of the $T^{\mu\nu}$ for
ideal fluid itself involves the conservation of the entropy (viscosity and
thermoconductivity, if any, are taken into account by including additional
terms in $T^{\mu\nu}$). Component perpendicular to the stream lines
results in the equation of stream lines equilibrium~--- stream equation
(or the Grad--Shafranov equation). With the terms
containing second derivatives written out explicitly it looks as follows
\begin{eqnarray}
&& D\left[\Delta\frac{\partial^2 \Phi}{\partial
r^2}+\sin\theta\frac{ \partial}{\partial
\theta}\left(\frac{1}{\sin\theta}\frac{\partial\Phi }{\partial\theta}
\right)\right]+\frac{1}{\Delta(\partial\Phi/\partial r)^2+(\partial\Phi
/\partial\theta)^2}\label{asc14}
\left[\Delta^2\left(\frac{\partial\Phi}{\partial r}\right)^2
\frac{\partial^2\Phi}{\partial r^2}+\right. \nonumber\\
&& \left.2\Delta\frac{\partial\Phi}{\partial r}
\frac{\partial\Phi}{\partial\theta}\frac{\partial^2\Phi}{\partial r
\partial\theta}+\left(\frac{\partial\Phi}{\partial\theta}\right)^2
\frac{\partial^2\Phi}{\partial\theta^2}-\frac{1}{2}\frac{\partial \Delta}{
\partial r}\frac{\partial \Phi}{\partial r}\left(\frac{\partial\Phi}{
\partial \theta}\right)^2\right]-
\frac{1}{2}\frac{\partial}{\partial r}\left(\frac{\rho^2}{\Delta}
\right)\frac{\Delta^2}{\rho^2}\frac{\partial\Phi}{\partial r}- \\
&& \frac{
1}{2}\frac{1}{\rho^2}\frac{\partial\rho^2}{\partial\theta}\frac{
\partial\Phi}{\partial\theta}-\frac{2\pi^2 n^2}{
|\mbox{\boldmath$\nabla$}\Phi|^2}\frac{\rho^2}{\mu^2}
\mbox{\boldmath$\nabla$}\Phi\cdot
\left[E^2\mbox{\boldmath$\nabla$}(\varpi^2)-\mu^2
\mbox{\boldmath$\nabla$}(\Delta\sin^2\theta)\right]=0\mbox{,} \nonumber
\end{eqnarray}
where
\begin{equation}
D=-1+\frac{(\Gamma-1)(\mu-m)}{m+(\mu-m)(2-\Gamma)}\frac{4\pi^2\Delta
\sin^2\theta n^2}{|\mbox{\boldmath$\nabla$}\Phi|^2}
=-1+\frac{1}{u_P^2}\frac{c_s^2}{1-c_s^2}\mbox{.}\label{asc15}
\end{equation}
Equation~(\ref{asc14}) with definition~(\ref{asc15}) and
relations~(\ref{asc8}) and~(\ref{asc13}) contains actually only flux
function~$\Phi(r,\theta)$, two constants~$s$ and~$E$ as well as metric
coefficients. It is elliptic  in the subsonic region~$D>0$ and hyperbolic
in the supersonic region~$D<0$. For the details of derivation of this
equation see the work by Beskin~\& Pariev~(1993). We need to find smooth
solution of the equations~(\ref{asc14}),~(\ref{asc15}),~(\ref{asc8})
and~(\ref{asc13}) for the function~$\Phi(r,\theta)$, considering boundary
conditions at infinity.

\section{Solution in the supersonic region}

Let us first find the solution in the region far under the sonic
surface $r=r_s$ for $r\ll r_s$, where we will treat it as being close to
the solution for radially infalling dust. In the case of cold matter
$k(s)$ is equal to~$0$, so $\mu=E=m$. By direct substitution into
equation~(\ref{asc14}), expressing~$n$ in terms of~$\Phi$ by means of
the relation~(\ref{asc13}), it is possible to convince yourself that any
arbitrary function~$\Phi=\Phi(\theta)$ is the solution. This just means
that the dust falling down to the black hole follows the
trajectory~$\theta=\mbox{const}$. At first, we need to specify this
arbitrary function in order to determine zero order approximation for
$\Phi$. In our case of nonrelativistic temperature of the
gas at infinity, sonic surface, at which the local sound velocity is equal
to the velocity of the flow, is situated far from the black hole horizon
$r_H={\cal M}+\sqrt{{\cal M}^2-a^2}$, in the region where relativistic
effects are weak.  Hence, the picture of the stream lines in the
transsonic region only slightly deviates from that for the accretion onto
Newtonian gravitating centre, and the main term in the expansion of~$\Phi$
corresponds to spherically symmetrical flow, that is proportional to
$1-\cos\theta$. We will show in section~4 that the corrections to
spherically symmetrical part of~$\Phi$ in transsonic region are of order of
the $({\cal M}/r_s)$, while the corrections to angular dependent part are
proportional to $({\cal M}/r_s)^2$ and $(a/r_s)^2$.

We define small parameter of the expansion~$\epsilon$ as the difference
between the entalpy per particle at infinity~$\mu_{\infty}$ and the rest
particle mass~$m$
\begin{equation}
\mu_{\infty}=m(1+\epsilon)\mbox{.}\label{asc16}
\end{equation}
Far from the sound surface and the black hole~${\bf u}=0$, hence, $E=\mu$.
This gives us exact expression for the constant~$E$
\begin{equation}
E=m(1+\epsilon)\mbox{.}\label{asc17}
\end{equation}
Introducing particle density at infinity~$n_{\infty}$, the expression for
entalpy~(\ref{asc8}) is rewritten in the form
\begin{equation}
\mu-m=\epsilon m\left(\frac{n}{n_{\infty}}\right)^{\Gamma-1}
\label{asc18}\mbox{.}
\end{equation}
It is well known that the accretion rate is determined by the condition of
passing the solution for~$n(r)$ smoothly through the sonic surface
(Shapiro~\& Teukolsky, 1983). Consideration of spherically symmetrical
Newtonian accretion problem~(e.g. Shapiro~\& Teukolsky, 1983) results in
the following accretion rate using our notations for entalpy~(\ref{asc16})
and~(\ref{asc18})
\begin{equation}
\Phi_0=\Phi_{00}\frac{1}{\epsilon^{3/2}}(1-\cos\theta)
\mbox{,}\label{asc19}
\end{equation}
where
\begin{equation}
\Phi_{00}=-2\pi n_{\infty}{\cal M}^2\varphi(\Gamma)\label{phi00}
\end{equation}
and
\begin{equation}
\varphi(\Gamma)=\left(\frac{1}{2}\right)^{\frac{\Gamma+1}{2(\Gamma-1)}}
\left(\frac{5-3\Gamma}{4}\right)^{-\frac{5-3\Gamma}{2(\Gamma-1)}}
\frac{1}{(\Gamma-1)^{3/2}}\mbox{.}\label{asc20}
\end{equation}
This expression is valid for~$1<\Gamma<5/3$. The case $\Gamma=5/3$ needs
special treatment and we will not concentrate on it though it
is this case that is closest to real description of infalling matter
if it is fully ionized. Thus, we have completely determined the
function~$\Phi_0(\theta)$, which was remained to be free in the case of
perfectly cold matter. We accept the expression~(\ref{asc19}) as a zero
order approximation to the solution. Note that, when $\epsilon\to 0$ and
$n_{\infty}=\mbox{const}$, the accretion rate will become infinitely
large. Further, by substitution the expression~(\ref{asc19}) into
equation~(\ref{asc13}) zero order approximation for particle
concentration~$n_0$ can be found
\begin{equation}
n_0(r)=n_{\infty}\varphi^2(\Gamma)\epsilon^{-3/2}
\frac{{\cal M}^2}{\sqrt{2{\cal M}r
(r^2+a^2)}}\mbox{.}\label{asc21}
\end{equation}
Also in zero order $\mu_0=m$, $D_0=-1$. Radius of the sonic surface is
\begin{equation}
r_{0s}={\cal M}\frac{1}{\epsilon}\frac{5-3\Gamma}{4(\Gamma-1)}\label{r0s}
\mbox{.}
\end{equation}
We see that $r_{0s}\gg {\cal M}$ for small~$\epsilon$ indeed.

The main equation of the problem~(\ref{asc14}) may be expanded
in parameter~$\epsilon$ only when all values, characterizing the flow,
deviate slightly from those for dust accretion. Therefore, matter needs
having nonrelativistic temperature not only at infinity but also in the
whole space. Let us check it. Substitution in expression~(\ref{asc18})
for~$n$ its zero order approximation~(\ref{asc21}) allows to find entalpy
with the first order accuracy
\begin{equation}
\mu-m=m\left(\varphi(\Gamma)\right)^{\Gamma-1}\epsilon^{(5-3\Gamma)/2}
\left(\frac{{\cal M}^3}{2r(r^2+a^2)}\right)^{(\Gamma-1)/2}
\mbox{.}\label{asc22}
\end{equation}
We see that if $\Gamma<5/3$, for $\epsilon\to 0$ and $r\sim {\cal M}$ the
correction to~$\mu$ is small indeed and the expansion in~$\epsilon$ is now
becoming full substantiated. For $\Gamma=5/3$ radius of the sonic
surface $\displaystyle r_{0s}={\cal M}\frac{3}{4}\sqrt{\frac{3}{2
\epsilon}}$ is still much larger than~${\cal M}$, but the temperature in
the vicinity of the event horizon is relativistic (for Schwarzschild black
hole $T_{H}\approx 0.111\,m$, Shapiro~\& Teukolsky~(1983)) and the
solution for dust accretion cannot be used as a zero order approximation.

Let us write for the flux function and particle density
$\Phi=\Phi_0+\Phi_1$ and $n=n_0+n_1$, where $\Phi_1\ll\Phi_0$ and $n_1\ll
n_0$, and perform the linearization of the equation~(\ref{asc14}) with
definition~(\ref{asc15}) and relation~(\ref{asc13}), allowing to
express~$n_1$ via~$\Phi_1$. Finally, after eliminating $n_1$ from
linearized equation~(\ref{asc14}) and doing some algebra, stream equation
can be reduced to the following simple equation in~$\Phi_1$
\begin{eqnarray}
&& \frac{\partial^2\Phi_1}{\partial r^2}+\frac{\partial\Phi_1}{\partial r}
\frac{3r^2+a^2}{2r(r^2+a^2)}=-\Phi_{00}\epsilon^{1-3\Gamma/2}\cos
\theta\sin^2\theta\times\label{asc23} \nonumber\\
&& \frac{a^2}{{\cal M}r(r^2+a^2)}\left(\varphi(\Gamma)\right)^
{\Gamma-1}\left(\frac{{\cal M}^3}{2r(r^2+a^2)}\right)^{(\Gamma-1)/2}
\mbox{.}
\end{eqnarray}
The remarkable fact is the cancellation of the derivatives~$\displaystyle
\frac{\partial^2 \Phi_1}{\partial\theta^2}$, $\displaystyle \frac{\partial
\Phi_1}{\partial\theta}$ and $\displaystyle \frac{\partial^2
\Phi_1}{\partial r\partial \theta}$, which makes
equation~(\ref{asc23}) to be an ordinary differential equation and enables
to find all its solutions.
Note, that the equation~(\ref{asc23}) does not
contain factor $\Delta$, which vanishes at the event horizon $r_H$. This
factor is cancelled out from all addenda of linearized
equation~(\ref{asc14}), hence, the equation~(\ref{asc23})  has no
singularities at the event horizon.  There is no even the characteristic
scale ${\cal M}$ in the latter equation, though from physical point of
view one can naturally expect it to be revealed. The only characteristic
size is $a$, which can be much smaller than~${\cal M}$.

We are able to find easily general solution of equation~(\ref{asc23}),
which can be written in the form such that the angular dependent part
of~$\Phi_1$ tends to~0 for $r\to\infty$
\begin{eqnarray}
&& \Phi_1=\Phi_{00}\,a^2\cos\theta\sin^2\theta\,
2^{-(\Gamma-1)/2}(\varphi(\Gamma))^{\Gamma-1} {\cal M}^{(3\Gamma-5)/2}
\epsilon^{1-3\Gamma/2}\times \nonumber\\
&& \int\limits_r^{+\infty}\frac{dr'}{\sqrt{r'({r'}^2
+a^2)}}\,\int\limits_{r_{H}}^{r'}\frac{dr''}{[r''({r''}^2+a^2)]^{\Gamma/2}}
\,-A_1(\theta)\int\limits_r^{+\infty}\frac{dr'}{\sqrt{r'({r'}^2+a^2)}}\,
+A_2(\theta)
\label{1}\mbox{,}
\end{eqnarray}
where $A_1$ and $A_2$ are arbitrary functions in $\theta$ only. In order
to match this solution with the solution in the transsonic region $r\sim
r_s$ obtained in section~4 let us expand expression~(\ref{1}) in $a/r\ll
1$
\begin{eqnarray}
&& \Phi_1=2\,\sqrt{\frac{{\cal M}}{r}}
\left(1-\frac{1}{10}\frac{a^2}{r^2}+
\frac{1}{24}\frac{a^4}{r^4}-\ldots \right)\left[\Phi_{00}C_0\left(
\frac{a}{{\cal M}}\right)^2\cos\theta\sin^2\theta\,
(\varphi(\Gamma))^{\Gamma-1}
\epsilon^{1-3\Gamma/2}\,2^{3/2-2\Gamma}\right. \nonumber\\
&& \biggl.-\frac{A_1(\theta)}{\sqrt{{\cal M}}}\biggr]- \label{2}
\Phi_{00}\left(\frac{a}{{\cal M}}\right)^2\cos\theta\sin^2\theta\, (\varphi
(\Gamma))^{\Gamma-1} 2^{-(\Gamma-1)/2}\epsilon^{1-3\Gamma/2}\,
\frac{4}{(3\Gamma-2)(3\Gamma-1)}\times \\
&& \left(\frac{{\cal M}}{r}\right)^{(3\Gamma-1)
/2}\left(1+\frac{3\Gamma-1}{3+3\Gamma}\frac{a^2}{r^2}-\frac{3}{8}\,
\frac{3\Gamma-1}{7+3\Gamma}\frac{a^4}{r^4}-\ldots\right)\,+A_2(\theta)
\mbox{,}       \nonumber
\end{eqnarray}
where constant $C_0$ denotes the value of the integral
$$ C_0=\int\limits_{r_{H}/(2{\cal M})}^{
+\infty}\left[\xi
\left(\xi^2+\frac{a^2}{
4{\cal M}^2}\right)\right]^{-\Gamma/2}d\xi\mbox{.}   $$
It is also useful to find the ratio of the perturbation $\Phi_1$ to
zero order approximation $\Phi_{00}\epsilon^{-3/2}$. Retaining only
leading terms in $a^2/r^2$ in expansion~(\ref{2}) one can obtain
\begin{eqnarray}
&& \frac{\Phi_1}{\Phi_{00}}\epsilon^{3/2}\approx2\,\sqrt{
\frac{{\cal M}}{r}}\left[
C_0\left(\frac{a}{{\cal M}}\right)^2\cos\theta\sin^2\theta\,
(\varphi(\Gamma))^{\Gamma-1}\,2^{3/2-2\Gamma}\,\epsilon^{(5-3\Gamma)/2}
-\frac{A_1(\theta)}{\Phi_{00}}\epsilon^{3/2}\right]- \label{3}
\nonumber\\
&& \left(\frac{a}{{\cal M}}\right)^2\cos\theta\sin^2\theta\,K(\Gamma)\left(
\frac{{\cal
M}}{r}\right)^{(3\Gamma-1)/2}\epsilon^{(5-3\Gamma)/2}\,+\frac{A_2
(\theta)}{\Phi_{00}}\epsilon^{3/2}\mbox{,}
\end{eqnarray}
where we have denoted for brevity
\begin{equation}
K(\Gamma)=(\varphi(\Gamma))^{\Gamma-1}\,2^{-(\Gamma-1)/2}
\frac{4}{(3\Gamma-        \label{K}
2)(3\Gamma-1)}\mbox{.}
\end{equation}
We cannot now determine the functions $A_1(\theta)$ and $A_2(\theta)$. It
will be done in section~4, where we obtain the solution of the stream
equation~(\ref{asc14}) in the region $r\sim {\cal M}/\epsilon$ and match
it to~(\ref{3}). At last, it is necessary to verify that $\Phi_1/\Phi_0\ll
1$.  Indeed, this is immediately seen from the equation~(\ref{3}) unless
functions $A_1$ and~$A_2$ have too high values, which is not the case as
it will be shown in section~4.

\section{Solution in the transsonic region}

Now let us consider accretion flow far from the event horizon but still on
the scales of order of the radius of the sound surface $r_s$. In this
region zero approximation for particle density~(\ref{asc21}) can no longer
be valid because when $r\sim {\cal M}/\epsilon$ main term in the
difference $\varpi^2E^2- \mu^2\Delta\sin^2\theta$ standing for particle
density in~(\ref{asc13}) $r^2E^2\sin^2\theta-\mu^2 r^2\sin^2\theta$
vanishes for cold matter, when $E=\mu=m$, so to determine this difference
correctly one must take into account thermal corrections to~$E$ and~$\mu$.
Clearly, the consideration of the lowest order corrections will lead us to
Bondi solution for the accretion in Newtonian gravitational potential of a
point mass. We have already known the flux function for that
solution~$\Phi_0$ given by formula~(\ref{asc19}). Therefore, zero order
approximations for $\Phi$ in supersonic and transsonic regions are one and
the same and are equal to~$\Phi_0$. However, zero order approximations for
entalpy and particle density are different.

To perform expansion of the equations~(\ref{asc13}), (\ref{asc18})
and~(\ref{asc14}) with definition~(\ref{asc15}) in small
parameter~$\epsilon$ in transsonic region, first, we shall introduce some
notations. Let us denote the ratio of the particle density to that at
infinity as $y$
$$ y(r,\theta)=\frac{n(r,\theta)}{n_{\infty}}\mbox{.} $$
Further it is necessary to introduce rescaled radius~$x$. We define it as
\begin{equation}
r={\cal M}\frac{x}{\epsilon}\label{r}
\end{equation}
Then, we will search for series in~$\epsilon$ for~$y$ and $\Phi$
\begin{equation}
y=y_0(x)+y_1(x,\theta)+y_2(x,\theta)+y_3(x,\theta)
+\ldots\label{yexp}\mbox{,}
\end{equation}
\begin{equation}
\Phi(r,\theta)=\Phi_{00}\epsilon^{-3/2}(1-\cos\theta+f_1(x,\theta)
+f_2(x,\theta)+f_3(x,\theta)+\ldots)\label{fexp}\mbox{,}
\end{equation}
where indexes $0,1,2,3,\ldots$ mark successive orders of expansion. For
$n$ and~$\Phi$ dependent on~$\theta$ sonic surface where $D=0$ will no
longer be spherical . Similar to the expansions of~$y$ and~$\Phi$  we
write an expansion of the radius of the sound surface~$r_s(\theta)$
\begin{equation}
x_s(\theta)=x_{0s}+x_{1s}(\theta)+x_{2s}(\theta)+x_{3s}(\theta)+\ldots
\label{sexp}\mbox{,}
\end{equation}
where the zero order value corresponds to $r_{0s}$ given by
the expression~(\ref{r0s})
\begin{equation}
x_{0s}=\frac{5-3\Gamma}{4(\Gamma-1)}\label{x0s}\mbox{.}
\end{equation}
Because an asymptotical expansion of metric coefficients $g_{ik}$ for
$r\to\infty$ contains all successive orders in~$\epsilon$ it is naturally
to expect that~$y_1$ and~$f_1$ will be of order of~$\epsilon$, $y_2$
and~$f_2$ of order of~$\epsilon^2$, $y_3$ and~$f_3$ of order
of~$\epsilon^3$, and so on. In further calculation dealing with series we
shall follow this order of values that finally will have occurred to be
true.
The only characteristic size in the transsonic region is~$r_{0s}$,
therefore, $\theta$- and~$x$- derivatives of the smooth functions~$y_0$,
$y_1$, $y_2$,~\ldots, and $f_1$, $f_2$,~\ldots are of the same order and
we shall regard in expansions $\displaystyle \frac{\partial f_1}{\partial
x}\sim \frac{\partial f_1}{\partial \theta}$,
$\displaystyle \frac{\partial y_2}{\partial
x}\sim \frac{\partial y_2}{\partial \theta}$, and so on.
In all subsequent expressions indexes $0,1,2,3,\ldots $ mark successive
orders of expansion of the values considered, index~$s$ means that the
quantity is taken at the nonperturbed sonic surface $x=x_{0s}$,
particularly, $\mu=\mu_0+\mu_1+\mu_2+\mu_3+\ldots $, $D=D_0+D_1+D_2+\ldots
$, and similarly for other values.

Now zero order approximation for entalpy is obtained by substituting
$y_0(x)$ for $n/n_{\infty}$ in the expression~(\ref{asc18})
\begin{equation}
\mu_0=m+\epsilon my_0^{\Gamma-1}\label{35}\mbox{.}
\end{equation}
Expansion in~$\epsilon$ of the relation~(\ref{asc13}) to the first order
using~(\ref{35}), (\ref{asc17}), and~(\ref{phi00}) results in the
following expression
\begin{equation}
1=y_0^{\Gamma-1}-\frac{1}{x}+\frac{\varphi^2(\Gamma)}{2y_0^2x^4}
\label{36}\mbox{,}
\end{equation}
which determines the function~$y_0(x)$. Differentiating~(\ref{36}) one can
obtain
\begin{equation}
\frac{dy_0}{dx}=\frac{y_0}{x}\frac{2\varphi^2(\Gamma)-x^3y_0^2}{(\Gamma-1)
y_0^{\Gamma+1}x^4-\varphi^2(\Gamma)}\label{37}\mbox{.}
\end{equation}
At the point $x=x_{0s}$, where the denominator in r.h.s. of~(\ref{37})
vanishes, the solution~$y_0(x)$ must be regular. This means that at
$x=x_{0s}$ the numerator must vanish too. Hence, we have three relations
among values~$x_{0s}$, $y_{0s}=y_0(x_{0s})$, and $\varphi(\Gamma)$; they
are equation~(\ref{36}), the equality to zero of the denominator
in~(\ref{37})
\begin{equation}
\varphi^2(\Gamma)=(\Gamma-1)y_{0s}^{\Gamma+1}x_{0s}^{4}\label{38}
\mbox{,}
\end{equation}
and the equality to zero of the numerator in~(\ref{37})
\begin{equation}
\varphi^2(\Gamma)=\frac{1}{2}x_{0s}^{3}y_{0s}^{2}\label{39}\mbox{.}
\end{equation}
Solution of these three relations gives the value of~$\varphi(\Gamma)$
just equal to that given by formula~(\ref{asc20}) used in section~3, the
value of~$x_{0s}$ equal to~(\ref{x0s}), and the following value
for~$y_{0s}$
\begin{equation}
y_{0s}=\left(\frac{2}{5-3\Gamma}\right)^{1/(\Gamma-1)}\label{40}
\mbox{.}
\end{equation}
With the relations~(\ref{38}), (\ref{39}) and~(\ref{40}) the
equation~(\ref{36}) determines two functions $y_0(x)$ passing smoothly
through the sonic point $x=x_{0s}$, one corresponding to Newtonian
accretion onto gravitating point object, the other to Newtonian ejection
from gravitating object~(Shapiro \& Teukolsky, 1983). Asymptotic of the
accretion solution for $x\to\infty$ is $\displaystyle y_0\to
1+\frac{1}{\Gamma-1}\frac{1}{x}$, of the ejection solution $y_0\sim
x^{-2}$; asymptotics for $x\to 0$ are: for the accretion solution
$\displaystyle y_0\to\frac{\varphi(\Gamma)}{\sqrt{2}}x^{-3/2}$, for the
ejection solution $y_0\sim x^{-1/(\Gamma-1)}$. We plotted functions
$y_0(x)$ on Fig.~1 for $\Gamma=4/3$. As it can be seen from the
asymptotical behaviour of these two functions their relative location for
all values $1<\Gamma<5/3$ is the same as on Fig.~1.  In order to determine
the derivative $\displaystyle \left.\frac{dy_0}{dx} \right|_{x=x_{0s}}$ we
expand the relation~(\ref{36}) up to the second order in the vicinity of
the point~$x=x_{0s}$, $y=y_{0s}$. Coefficients in the first order
expansion identically vanish due to the conditions~(\ref{38})
and~(\ref{39}). As a result, we obtain quadratic equation in the
derivative sought, which has two solutions, one corresponding to
the accretion curve~$y_0(x)$, the other to the ejection curve of Bondi
solution.  Note that no additional conditions should be imposed in order
to ensure the existence and continuity of all higher derivatives of
$y_0(x)$ at the sonic point. Because r.h.s. of the equation~(\ref{36}) is
function analytic  in both~$x$ and~$y_0$ in the neighbourhood of
$x=x_{0s}$ two functions~$y_0(x)$ determined implicitly by the
equation~(\ref{36}) are analytic too.  Since the lines $y_0=y_0(x)$ are
lines of constant value of the r.h.s. of the equation~(\ref{36}), they
cannot intersect each other save the sonic point. Therefore, deducing from
the asymptotical behaviour of the accretion and ejection solutions, one
must choose for the derivative of $y_0(x)$ at the sonic point larger value
from two. Thus, we obtain
\begin{equation}
\left.\frac{dy_0}{dx}\right|_{x=x_{0s}}=2\frac{\Gamma-1}{\Gamma+1}
\left(\frac{2}{5-3\Gamma}\right)^{\Gamma/(\Gamma-1)}\left(-4+\sqrt{10-
6\Gamma}\right)\label{42}\mbox{.}
\end{equation}

Zero order of~$D$ given by the formula~(\ref{asc15}) is
\begin{equation}
D_0=-1+\frac{\Gamma-1}{\varphi^2(\Gamma)}y_0^{\Gamma+1}x^4\label{43}
\mbox{.}
\end{equation}
At the sonic point $D_0(x_{0s})=0$.
In the stream equation~(\ref{asc14})
$$ L_{\theta}\Phi_0=\sin\theta\frac{\partial}{\partial\theta}\left(
\frac{1}{\sin\theta}\frac{\partial\Phi_0}{\partial\theta}\right)=0 $$
and substitution of zero order values in all other terms of
equation~(\ref{asc14}) leads after cancelling out the factor $\cos\theta$
to the relation~(\ref{36}). Hence, in the zero order approximation stream
equation~(\ref{asc14}) is satisfied automatically.

\subsection{First order corrections}

Now let us turn to the first order corrections. From~(\ref{asc18}) we
obtain
\begin{equation}
\mu_1=\epsilon m y_0^{\Gamma-1}\frac{y_1}{y_0}(\Gamma-1)\label{44}
\mbox{.}
\end{equation}
Expansion of the quantity $\displaystyle \frac{4\pi^2 n^2}{|
\mbox{\boldmath$\nabla$}\Phi|^2}$ entering both the relation~(\ref{asc13})
and the equation~(\ref{asc14}) up to the first order is
\begin{equation}
\left. \frac{4\pi^2 n^2}{|\mbox{\boldmath$\nabla$}\Phi|^2}\right|_0\,
+\>\left.\frac{4\pi^2 n^2}{|\mbox{\boldmath$\nabla$}\Phi|^2}\right|_1 \,
=\> \frac{\epsilon x^2y_0^2}{{\cal M}^2\varphi^2(\Gamma)\sin^2\theta}
\left(1+2\frac{y_1}{y_0}-\frac{2}{\sin\theta}\frac{\partial f_1}{
\partial\theta}\right)\label{45}\mbox{.}
\end{equation}
Because of the vanishing of the leading term $\sim 1/\epsilon^2$ in the
relation~(\ref{asc13}) we need to expand it to the second order and,
hence, one might expect that the rotational parameter~$a$ will be involved
in the first order approximation, because it enters quadratically into
the expressions~(\ref{asc5}) for~$\varpi^2$ and~$\Delta$. However,
calculation shows that the terms containing~$a^2$ are cancelled out from
the both sides of the equation~(\ref{asc13}), thus resulting in the
following expression for~$y_1$ through~$f_1$ \begin{equation}
2\frac{y_1}{y_0}\left((\Gamma-1)y_0^{\Gamma-1}-\frac{\varphi^2(\Gamma)}{
x^4y_0^2}\right)=\epsilon(1+3y_0^{2\Gamma-2}-4y_0^{\Gamma-1})-
\frac{\varphi^2(\Gamma)}{x^4y_0^2}\frac{2}{\sin\theta}\frac{\partial f_1}{
\partial\theta}\label{46}\mbox{.}
\end{equation}
Further, we have to linearize equation~(\ref{asc14}) using~(\ref{44})
and~(\ref{45}) and insert into it the expression~(\ref{46}) for~$y_1$. The
$\theta$--component of the expression embraced by the square brackets in
the last term of l.h.s. of~(\ref{asc14}) $\displaystyle \> \frac{E^2}{\mu^2
}\frac{\partial\varpi^2}{\partial\theta}-2\Delta\sin\theta\cos\theta\>$
must be expanded to the second order (because of the vanishing leading
term in this difference) and, again, $a^2$ is cancelled, and the
linearization of the stream equation does not contain~$a^2$. For~$D$ it is
enough to use its zero order~$D_0$; in the linearized last term in l.h.s.
of~(\ref{asc14}) $y_1$ stands only multiplied by the factor
$\displaystyle \>
\frac{2}{y_0}\left((\Gamma-1)y_0^{\Gamma-1}-\frac{\varphi^2(\Gamma)}{
x^4y_0^2}\right)\>$ that allows it to be directly replaced by the r.h.s.
of expression~(\ref{46}). This replacement results in the cancellation of
the free term in stream equation. Finally, dividing~(\ref{asc14}) by
$\Phi_{00}\epsilon^{-3/2}$ we obtain the following linear second order
differential equation in~$f_1$
\begin{equation}
D_0x^2\frac{\partial^2 f_1}{\partial
x^2}+(D_0+1)\sin\theta\frac{\partial}{\partial\theta}\left(\frac{1}{
\sin\theta}\frac{\partial f_1}{\partial\theta}\right)-
\left(2x-\frac{x^4y_0^2}{\varphi^2(\Gamma)}\right)\frac{\partial f_1}{
\partial x}=0\label{47}\mbox{,}
\end{equation}
where~$D_0$ is given by the expression~(\ref{43}).

Equation~(\ref{47}) is an equation with separable variables and angular
operator
\begin{equation}
\displaystyle L_{\theta}=
\sin\theta\frac{\partial}{\partial\theta}\left(\frac{1}{
\sin\theta}\frac{\partial }{\partial\theta}\right)
\end{equation}
is the same as in the
linearizations of the stream equation considered in BP and by Petrich,
Shapiro~\& Teukolsky~(1988). Particle flux density must be finite at the
rotational axis. As it is seen from the expression~(\ref{asc6}) for $n{\bf
u}_P$ this leads to that the quantity $\displaystyle
\frac{1}{\sin\theta}\frac{\partial f_1}{\partial\theta}$ must be finite at
$\theta=0$ and $\theta=\pi$. Besides, it is obvious that at $\theta=0$
total flux must be equal to~0, so one more boundary condition for~$f_1$ is
$\displaystyle f_1|_{\theta=0}=0$. It can be proved that the eigenvalue
problem for operator $L_{\theta}$ with these boundary conditions has
eigenvalues $C_n=-n(n+1)$, where $n=0,\,1,\,2,\,3,\ldots$, with
corresponding eigenfunctions
$$ Q_n=\frac{2^n n!(n-1)!}{(2n)!}\sin^2\theta P^{'}_n(\cos\theta)
\mbox{,}$$
where prime denotes differentiation of the Legendre polynomial $P_n(\xi)$
with respect to~$\xi$ and normalizing factor is chosen the same as in~BP
to make the comparison of our results to the results by~BP more direct.
First four eigenfunctions are
\begin{equation}
Q_0=1-\cos\theta\mbox{,}\quad Q_1=\sin^2\theta\mbox{,} \quad
Q_2=\sin^2\theta\cos\theta\mbox{,} \quad Q_3=\sin^2\theta\left(
\cos^2\theta-\frac{1}{5}\right)\label{49}\mbox{.}
\end{equation}
We look for the solution of the equation~(\ref{47}) in the form
\begin{equation}
f_1=\sum\limits_{n=0}^{\infty}\epsilon g_{1n}(x)Q_n(\theta)
\label{50}\mbox{.}
\end{equation}
Then $g_{1n}(x)$ must satisfy the following equation
\begin{equation}
D_0 x^2\frac{d^2g_{1n}}{dx^2}+\left(\frac{x^4y_0^2}{\varphi^2(\Gamma)}-2x
\right)\frac{dg_{1n}}{dx}-(D_0+1)n(n+1)g_{1n}=0\label{51}
\end{equation}
with boundary conditions that $g_{1n}(x)$ must be regular at the sonic
point where $D_0=0$ and be finite at $x\to\infty$. Two linearly independent
solutions of~(\ref{51}) have asymptotics $\sim x^{n+1}$ and $\sim x^{-n}$,
so we must choose the latter solution. Because of the relation~(\ref{39})
it follows from equation~(\ref{51}) that $n(n+1)g_{1n}(x_{0s})=0$. For
$n\neq 0$ it means that $g_{1n}(x)\equiv 0$, while for $n=0$ one obvious
solution of uniform equation~(\ref{51}) is $g_{10}=\mbox{const}$. Second
linearly independent solution has asymptotic $g_{10}\sim x$ for
$x\to\infty$ and must be ruled out. Thus, first order correction to the
flux function $f_1$ has the form
\begin{equation}
f_1=\epsilon g_{10}(1-\cos\theta)\label{52}\mbox{.}
\end{equation}

To complete construction of the first order approximation it remains to
calculate~$y_1$ according to the equation~(\ref{46}). We see that~$y_1$
does not depend on~$\theta$. Correction to particle density~$y_1(x)$ must
be smooth function in the vicinity of the sonic point $x=x_{0s}$, where
according to~(\ref{38}) l.h.s. of the equation~(\ref{46}) vanishes.
Therefore, r.h.s. must vanish too, which allows us to determine the value
of~$g_{10}$
\begin{equation}
g_{10}=\frac{3}{4}\,\frac{3\Gamma+1}{5-3\Gamma}\mbox{.}\label{53}
\end{equation}
Now $y_1$ is given by the formula
\begin{equation}
y_1=\epsilon y_0\,\frac{1+3y_0^{2\Gamma-2}-4y_0^{\Gamma-1}-2g_{10}
\frac{\varphi^2(\Gamma)}{x^4y_0^2}}{2(\Gamma-1)y_0^{\Gamma-1}-2
\frac{\varphi^2(\Gamma)}{x^4y_0^2}}\label{54}\mbox{.}
\end{equation}
Both numerator and denominator in~(\ref{54}) are analytic functions
in~$y_0$ and~$x$, therefore, $y_1(x)$ is also analytic function in the
neighbourhood of the sonic point, and no additional constrains should be
imposed in order of continuity of the derivatives of~$y_1$. The value
of~$y_1$ at the sonic point is
\begin{equation}
y_1|_{x=x_{0s}}=\,\frac{\epsilon y_{0s}}{2(5-3\Gamma)\sqrt{10-6\Gamma}}
\,\left[\frac{21\Gamma-5}{\Gamma+1}\left(-4+\sqrt{10-6\Gamma}\right)
+6(3\Gamma+1)\right]\label{55}\mbox{.}
\end{equation}
Asymptotical behaviour for $x\to\infty$ is $\displaystyle  y_1\to
\frac{\epsilon}{x(\Gamma-1)}$. Function~$y_1(x)$ is plotted on Fig.~2 for
some values of~$\Gamma$.

What we actually have obtained considering first order approximation are
the corrections to Newtonian accretion rate and particle density in the
problem of spherically symmetrical accretion onto a Schwarzschild black
hole. As long as~$a$ is not contained in the results they coincide with
those for a Schwarzschild black hole. Indeed, exact analytical solution
for the accretion onto a Schwarzschild black hole (e.g., Shapiro~\&
Teukolsky, 1983) in our notations can be written as
$$
\Phi=\Phi_{sw}(1-\cos\theta)
$$
with
\begin{equation}
\Phi_{sw}=-2\pi n_{\infty}{\cal M}^2\epsilon^{-1/(\Gamma-1)}\frac{1}{4}
\left(\frac{{\cal M}}{2r_s}\right)^{\frac{5-3\Gamma}{2(\Gamma-1)}}
\frac{1}{\left(\Gamma-1+\frac{{\cal M}}{2r_s}(2-3\Gamma)\right)^{
1/(\Gamma-1)}}\label{56}\mbox{,}
\end{equation}
where the value $\displaystyle \frac{{\cal M}}{2r_s}\sim \epsilon$ should
be obtained by solving cubic equation
\begin{equation}
(1+\epsilon)^2\left(1-\frac{3\Gamma-2}{\Gamma-1}\,\frac{{\cal M}}{2r_s}
\right)^2=\left(1-3\frac{{\cal M}}{2r_s}\right)^3
\label{57}\mbox{.}
\end{equation}
For the rescaled radius of the sonic surface $x_{sw}=\epsilon r_s/{\cal
M}$ expansion of the solution of equation~(\ref{57}) in terms
of~$\epsilon$ up to $\epsilon^3$ is
\begin{eqnarray}
&& x_{sw}=\,\frac{5-3\Gamma}{4(\Gamma-1)}\,\left[1+\epsilon
\frac{30\Gamma-9\Gamma^2-13}{2(5-3\Gamma)^2}+\epsilon^2\frac{27}{4}\,
\frac{(3\Gamma^2-14\Gamma+3)(\Gamma-1)^2}{(5-3\Gamma)^4}- \right.
\nonumber\\
&& \left.\epsilon^3
\frac{27(\Gamma-1)^2}{8(5-3\Gamma)^6}(3\Gamma+1)(9\Gamma^3-75\Gamma^2
+103\Gamma-53)\right]\label{58}\mbox{.}
\end{eqnarray}
For all $1<\Gamma<5/3$  the successive terms in the
series~(\ref{58}) have alternating signs: $x_{sw1}>0$, $x_{sw2}<0$,
$x_{sw3}>0$. Substitution of the expression~(\ref{58}) into
formula~(\ref{56}) results in series for the accretion rate
\begin{eqnarray}
&& \Phi_{sw}=\Phi_{00}\epsilon^{-3/2}\left(1+\epsilon\frac{3}{4}\,
\frac{3\Gamma+1}{5-3\Gamma}-\epsilon^2\frac{3}{32}\,\frac{135\Gamma^3-
99\Gamma^2-207\Gamma+139}{(5-3\Gamma)^3}+\right. \label{59}\nonumber\\
&& \left.\epsilon^3\,\frac{8505\Gamma^5-9153\Gamma^4-37530\Gamma^3+
89226\Gamma^2-71583\Gamma+20279}{128(5-3\Gamma)^5}\right)\mbox{.}
\end{eqnarray}
First order term in~(\ref{59}) coincides with the expression for
$g_{10}$~(\ref{53}) and is positive, which means enhanced accretion rate.
Term proportional to~$\epsilon^2$ is positive for $1<\Gamma<1.3037$ and
negative for $1.3037<\Gamma<5/3$, term proportional to~$\epsilon^3$ is
positive for $1.1558<\Gamma<5/3$ and negative for $1<\Gamma<1.1558$.

At last, let us show that the correction to the radius of the sonic
surface~$x_{1s}$ (see expansion~(\ref{sexp})) does not depend on~$\theta$
and is equal to~$x_{sw1}$ from~(\ref{58}). In general sonic surface is
located where~$D=0$. First order of~$D$ given by the formula~(\ref{asc15})
is
\begin{equation}
D_1=(\Gamma-1)\frac{y_0^{\Gamma+1}x^4}{\varphi^2(\Gamma)}\left((\Gamma+1)
\frac{y_1}{y_0}-(2-\Gamma)\epsilon y_0^{\Gamma-1}-\frac{2\epsilon}{x}-
2\epsilon g_{10}\right)\label{60}
\end{equation}
and does not depend on~$\theta$. Correction~$x_{1s}$ is determined from
the equation
$$
\left.\frac{dD_0}{dx}\right|_{x=x_{0s}}x_{1s}+D_1(x_{0s})=0\mbox{.}
$$
Its solution is readily obtained using expressions~(\ref{42})
and~(\ref{55})
\begin{equation}
x_{1s}=\epsilon\,\frac{30\Gamma-9\Gamma^2-13}{8(\Gamma-1)(5-3\Gamma)}
\label{61}
\end{equation}
and coincides with~$x_{sw1}$.

\subsection{Second order corrections}

In order to match the solution in the supersonic region we need to
consider further orders of expansion in the transsonic region.
Finding~$y_2$ and~$f_2$ we will act similarly as when we deal with~$y_1$
and~$f_1$. Let us denote by $\hat a=a/{\cal M}$ dimensionless value
of~$a$.  From~(\ref{asc18}) we obtain $$ \mu_2=\epsilon m
y_0^{\Gamma-1}(\Gamma-1)\left(\frac{y_2}{y_0}+
\frac{\Gamma-2}{2}\frac{y_1^2}{y_0^2}\right)\mbox{,} $$
further,
$$
\left.\frac{4\pi^2 n^2}{|\mbox{\boldmath$\nabla$}\Phi|^2}\right|_2=
\frac{\epsilon x^2}{{\cal M}^2\varphi^2(\Gamma)\sin^2\theta}\left(
2y_0 y_2+y_1^2+\epsilon^2\frac{{\hat a}^2 }{x^2}\cos^2\theta y_0^2-
4\epsilon g_{10}y_0 y_1-\frac{2}{\sin\theta}\frac{\partial f_2}{\partial
\theta}y_0^2+3\epsilon^2 g_{10}^2 y_0^2\right)\mbox{.} $$
Third order of expansion of relation~(\ref{asc13}) gives an expression
for~$y_2$ as a function of unknown~$f_2$
\begin{eqnarray}
&& 2\frac{y_2}{y_0}\left(y_0^{\Gamma-1}(\Gamma-1)-\frac{\varphi^2(\Gamma)
}{x^4y_0^2}\right)=\epsilon^2\left[\frac{{\hat a}^2}{x^2}\frac{
\varphi^2(\Gamma)}{y_0^2x^4}+2\frac{{\hat a}^2}{x^2}\cos^2\theta (1-
y_0^{\Gamma-1})+2y_0^{2\Gamma-2}(y_0^{\Gamma-1}-1) \right. \nonumber\\
&& \left.-g_{10}(g_{10}+4y_0^{\Gamma   \label{62}
-1})\frac{\varphi^2(\Gamma)}{y_0^2 x^4}\right]+2\epsilon\frac{y_1
}{y_0}\left[y_0^{2\Gamma-2}(\Gamma-1)-2(\Gamma-1)y_0^{\Gamma-1}+
2y_0^{\Gamma-1}\frac{\varphi^2(\Gamma)}{y_0^2x^4}\right. \\
&& \left.+2g_{10}
\frac{\varphi^2(\Gamma)}{y_0^2 x^4}\right]-
\frac{y_1^2}{y_0^2}\left[(\Gamma-1)(\Gamma-2)y_0^{\Gamma-1}+3
\frac{\varphi^2(\Gamma)}{y_0^2 x^4}\right]-\frac{2}{\sin\theta}\frac{
\partial f_2}{\partial\theta}\frac{\varphi^2(\Gamma)}{y_0^2 x^4}\mbox{.}
\nonumber
\end{eqnarray}
Note that when obtaining second order of the stream equation~(\ref{asc14})
we need to retain only leading term~$D_0$ in the series for~$D$ because
both~$\Phi_0$ and~$\Phi_1$ do not depend on~$r$ and $L_{\theta}\Phi_0=
L_{\theta}\Phi_1=0$. Therefore, partial differential equation for~$f_2$
changes its type at the surface~$x=x_{0s}$ as well as equation~(\ref{47})
for~$f_1$ does. Again, $y_2$ appears in the second order of the stream
equation only multiplied by $\displaystyle
\frac{2}{y_0}\left(y_0^{\Gamma-1}(\Gamma-1)-\frac{\varphi^2(\Gamma)
}{x^4y_0^2}\right)$, which allows it to be excluded by means of
expression~(\ref{62}). Finally, dividing~(\ref{asc14}) by
$\Phi_{00}\epsilon^{-3/2}$ one can obtain the following linear second
order partial differential equation in~$f_2$
\begin{eqnarray}
&& D_0 x^2\frac{\partial^2 f_2}{\partial x^2}+(D_0+1)\sin\theta\frac{
\partial}{\partial\theta}\left(\frac{1}{\sin\theta}\frac{\partial f_2}{
\partial\theta}\right)-\left(2x-\frac{y_0^2x^4}{\varphi^2(\Gamma)}
\right)\frac{\partial f_2}{\partial x}+ \label{63}\nonumber\\
&& \epsilon^2\frac{{\hat a}^2}{x^2}\cos\theta\sin^2\theta\left(
1-\frac{2}{x}\frac{y_0^2x^4}{\varphi^2(\Gamma)}\right)=0\mbox{.}
\end{eqnarray}
Variables in equation~(\ref{63}) can be separated and angular operator
coincides with that in equation~(\ref{47}). Moreover, equation~(\ref{63})
differs from equation~(\ref{47}) only by that the former contains free
term, while the latter does not. Equation~(\ref{63}) coincides with the
limit~$c_s^2\ll$~1 of the equation~(52) in~BP, which is the linearization
of the stream equation in small parameter~$a/{\cal M}$, so the results
of~BP concerning the case $c_s^2\ll 1$ can be applied to the
equation~(\ref{63}). Similar to the equation~(\ref{47}) we seek for the
solution in the form
$$ f_2=\sum\limits_{n=0}^{\infty}\epsilon^2g_{2n}(x)Q_n(\theta)
\mbox{.}$$
Then, radial functions $g_{2n}(x)$ for all $n\neq 2$ must satisfy exactly
the same equation as~(\ref{51}) with identical boundary conditions. This
immediately allows to conclude that $g_{2n}\equiv 0$ for all $n\neq
0,\,2$ and $g_{20}=\mbox{const}$. As for~$g_{22}$ it must satisfy the
following equation
\begin{equation}
-D_0\frac{d^2 g_{22}}{dx^2}+\left(\frac{2}{x}-\frac{1}{x^2}\frac{y_0^2
x^4}{\varphi^2(\Gamma)}\right)\frac{dg_{22}}{dx}+\frac{6}{x^2}
(D_0+1)g_{22}=\frac{{\hat a}^2}{x^4}\left(1-\frac{2}{x}\frac{y_0^2x^4}{
\varphi^2(\Gamma)}\right)\label{64}\mbox{,}
\end{equation}
which is identical to equation~(57) in~BP. The requirement of regularity
of~$g_{22}(x)$ at nonperturbed sound surface $D_0=0$ leads to the
condition
\begin{equation}
g_{22}(x_{0s})=-\frac{{\hat a}^2}{2x_{0s}^2}\label{65}\mbox{.}
\end{equation}
Another boundary condition for~$g_{22}$ is $g_{22}\sim x^{-2}$ when
$x\to\infty$. Similarly to~BP we introduce function~$G(x)$ such that the
solution of~(\ref{64}) satisfying boundary conditions is written in the
form
$$ g_{22}(x)=-G(x)\frac{{\hat a}^2}{x_{0s}^2}\left(\frac{x}{x_{0s}}
\right)^{(1-3\Gamma)/2}\mbox{,} $$
where $G(x_{0s})=1/2$ and $G(x)\sim x^{(3\Gamma-5)/2}$ for $x\to\infty$.
The dependencies~$G(x)$ found numerically by integrating~(\ref{64}) with
boundary conditions are shown on Fig.~3, which is actually the same as
Fig.~1 in~BP. Thus, second order correction to the flux function~$f_2$ has
the form
\begin{equation}
f_2=\epsilon^2 g_{20}(1-\cos\theta)-\epsilon^2G(x)\frac{{\hat a}^2}{
x_{0s}^2}\left(\frac{x}{x_{0s}}\right)^{(1-3\Gamma)/2}\sin^2\theta\cos
\theta\label{66}\mbox{.}
\end{equation}

Similar to constant~$g_{10}$ constant $g_{20}$ should be determined  from
the requirement of smoothness of the value~$y_1$ at the sonic surface
$x=x_{0s}$. Inserting into expression~(\ref{62}) value for~$f_2$
from~(\ref{66}) we see that
$$ \frac{1}{\sin\theta}\frac{\partial f_2}{\partial \theta}=\epsilon^2
g_{20}+\epsilon^2 g_{22}(x)(3\cos^2\theta-1)\mbox{.} $$
Therefore, expression~(\ref{62}) contains terms proportional to
$\cos^2\theta$ and terms dependent only on~$x$. For evaluating r.h.s.
of~(\ref{62}) at the nonperturbed sonic surface we use
expressions~(\ref{65}) for~$g_{22}(x_{0s})$, (\ref{55}) for~$y_1(x_{0s})$,
(\ref{40}) for~$y_0(x_{0s})$, and~(\ref{53}) for~$g_{10}$. As a result,
terms proportional to $\cos^2\theta$ vanish, while the requirement for
the rest addenda to be equal to~0 determines the value of~$g_{20}$
\begin{equation}
g_{20}=-\frac{3}{32}\,\frac{135\Gamma^3-
99\Gamma^2-207\Gamma+139}{(5-3\Gamma)^3}\label{g20}\mbox{,}
\end{equation}
which actually coincides with the second order term in the expansion
of~$\Phi_{sw}$~(\ref{59}). This result is obvious because in the case
$a^2=0$ when $g_{22}=0$ and~$f_2=\epsilon^2 g_{20}(1-\cos\theta)$ we deal
with the accretion onto a Schwarzschild black hole. Now the construction
of the second order approximation is completed.

Now we are able to match the solution in the supersonic region obtained in
section~3 with the solution in the transsonic region. For this purpose we
should consider the behaviour of~$g_{22}(x)$ for $x\to 0$. Expanding the
relation~(\ref{36}) in the vicinity of~$x=0$ one can find two leading
terms in the asymptotical series for~$y_0(x)$
$$ y_0=\frac{\varphi(\Gamma)}{\sqrt{2}}x^{-3/2}\left[1+\frac{(\varphi
(\Gamma))^{\Gamma-1}}{2^{(\Gamma+1)/2}}x^{(5-3\Gamma)/2}+o\left(x^{
(5-3\Gamma)/2}\right)\right]\mbox{.} $$
Substitution of this expression into r.h.s. of~(\ref{64}) allows us to
find first nonzero term in the expansion of r.h.s. of~(\ref{64}). Hence,
in the limit $x\to 0$ equation~(\ref{64}) takes the form
$$ \frac{d^2g_{22}}{dx^2}+\frac{3}{2x}\frac{dg_{22}}{dx}=-{\hat a}^2
\frac{(\varphi(\Gamma))^{\Gamma-1}}{2^{(\Gamma-1)/2}}x^{-3(\Gamma+1)/2}
\mbox{.}  $$
General solution of the last equation is
\begin{equation}
g_{22}=C_1 x^{-1/2}+C_2-{\hat a}^2x^{(1-3\Gamma)/2}K(\Gamma)\label{67}
\mbox{,}
\end{equation}
where the constant~$K(\Gamma)$ occurs to be just equal to that given
by~(\ref{K}) and constants~$C_1$ and~$C_2$ are uniquely determined by
boundary conditions for $g_{22}$. Since~$g_{22}(x)$ satisfies boundary
condition~(\ref{65}) at the sonic point, constants~$C_1$ and~$C_2$ are of
order of~${\hat a}^2$ and it is convenient to write them explicitly
$C_1=c_1(\Gamma){\hat a}^2$, $C_2=c_2(\Gamma){\hat a}^2$.
Therefore, in the inner region $x\sim\epsilon$ the
third term in the expression~(\ref{67}) is dominant. This conclusion is
confirmed by numerical computation of~$g_{22}(x)$, the results of which
are plotted on Fig.~3. At $x\to 0$ function~$G(x)$ tends to a constant
value, which demonstrates that the first two terms in the
expression~(\ref{67}) become negligible compared to the third one.
Limiting value of~$G(x)$ is derived from the last term of~(\ref{67})
$$ \lim_{x\to 0}G(x)=\frac{4}{(3\Gamma-2)(3\Gamma-1)(\Gamma-1)2^{\Gamma}}
$$
and is in good agreement with the numerical results. Limit $r\gg {\cal
M}$ of the inner solution~(\ref{3}) when expressed in terms of rescaled
variable~$x$ takes the form
\begin{eqnarray}
&& \left.\frac{\Phi_1}{\Phi_{00}}\epsilon^{3/2}\right|_{\rm in}=\>2C_0
{\hat a}^2\cos\theta\sin^2\theta(\varphi(\Gamma))^{\Gamma-1}2^{3/2-
2\Gamma}x^{-1/2}\epsilon^{3(2-\Gamma)
/2}-2\frac{A_1(\theta)}{\Phi_{00}\sqrt{{\cal M}}}x^{-1/2}
\epsilon^2- \nonumber\\
&& {\hat a}^2\cos\theta\sin^2\theta K(\Gamma)x^{(1-3\Gamma)/2}\epsilon^2+
\frac{A_2(\theta)}{\Phi_{00}}\epsilon^{3/2}\label{68}\mbox{.}
\end{eqnarray}
For outer solution~(\ref{fexp}) combining~(\ref{52}),~(\ref{66}) we have
\begin{eqnarray}
&& \left.\frac{\Phi_1}{\Phi_{00}}\epsilon^{3/2}\right|_{\rm out}=\>
\epsilon^2C_1\cos\theta\sin^2\theta x^{-1/2}+\epsilon^2C_2\cos\theta\sin^2
\theta-\epsilon^2{\hat a}^2x^{(1-3\Gamma)/2}K(\Gamma)\cos\theta\sin^2
\theta \nonumber\\
&& \label{69} +(1-\cos\theta)(\epsilon g_{10}+\epsilon^2 g_{20})+O(
\epsilon^3)\mbox{,}
\end{eqnarray}
where symbol~$O(\epsilon^{\alpha})$ denotes terms proportional
to~$\epsilon^{\alpha}$. In order to match expression~(\ref{68})
to~(\ref{69}) we must choose the following values for
functions~$A_1(\theta)$ and~$A_2(\theta)$
\begin{eqnarray}
&& A_1(\theta)=\Phi_{00}\sqrt{{\cal M}}{\hat a}^2\cos\theta\sin^2\theta
\left[C_0(\varphi(\Gamma))^{\Gamma-1}2^{3/2-2\Gamma}\epsilon^{1-3\Gamma/2}
-\frac{1}{2}c_1(\Gamma)\right]+O(\epsilon)\mbox{,} \nonumber\\
&& A_2(\theta)=\Phi_{00}{\hat a}^2 \epsilon^{1/2}c_2(\Gamma)\cos\theta
\sin^2\theta+\Phi_{00}(1-\cos\theta)\epsilon^{-1/2}(g_{10}+\epsilon
g_{20})+O(\epsilon^{3/2})\mbox{.} \nonumber
\end{eqnarray}
With these expressions solution~(\ref{1}) in the supersonic region becomes
fully determined and can be expressed by the following formula
\begin{eqnarray}
&& \frac{\Phi_1}{\Phi_{00}}\epsilon^{3/2}=-\frac{a^2}{{\cal M}^2}\cos\theta
\sin^2\theta(\varphi(\Gamma))^{\Gamma-1}2^{1-2\Gamma}\epsilon^{(5-3
\Gamma)/2} \int\limits_{r/(2{\cal M})}^{+\infty}\frac{d\xi^{'}}{
\sqrt{\xi^{'}(\xi^{'2}+{\hat a}^2/4)}}\times \nonumber \label{70} \\
&& \int\limits_{\xi^{'}}^{+\infty}
\frac{d\xi^{''}}{[\xi^{''}(\xi^{''2}+{\hat a}^2/4)]^{\Gamma/2}}
+\frac{a^2}{{\cal M}^2}\cos\theta\sin^2\theta\frac{c_1(\Gamma)}{2^
{3/2}}\epsilon^{3/2}\int\limits_{r/(2{\cal M})}^{+\infty}\frac{d\xi^{'}}{
\sqrt{\xi^{'}(\xi^{'2}+{\hat a}^2/4)}}+ \\
&& \epsilon^2\frac{a^2}{{\cal M}^2}c_2(\Gamma)\cos\theta\sin^2\theta+
(1-\cos\theta)(\epsilon g_{10}+\epsilon^2 g_{20})+O(\epsilon^{5/2})
\mbox{.} \nonumber
\end{eqnarray}

Firstly, we see that $\Phi_1\ll \Phi_0$. Secondly, dominating term
in~(\ref{70}) is the first one and, since we deal only with the first
order approximation to~$\Phi$, we cannot be sure that higher order terms
are negligible in comparison with the rest addenda in~(\ref{70}). At the
same time, only terms dependent on~$\theta\>$ like $Q_0=1-\cos\theta\>$
contribute to the total accretion rate, and because the total accretion
rate must be some constant value independent of the radius these terms do
not depend on~$r$ in all orders in~$\epsilon$. Hence, these terms are the
same in the supersonic region and in the transsonic region and are
determined correctly  in the
formula~(\ref{70}) up to the $O(\epsilon^3)$.
Finally, we write down explicitly the solution in
the transsonic region which matches the solution~(\ref{70})
\begin{equation}
\Phi=\Phi_{00}\epsilon^{-3/2}[1+(\epsilon g_{10}+\epsilon^2 g_{20})(1-
\cos\theta)+\epsilon^2 g_{22}(r)\cos\theta\sin^2\theta+O(\epsilon^3)]
\label{71}\mbox{.}
\end{equation}
On Fig.~4 we plotted the picture of the stream lines corresponding to the
flux function given by formula~(\ref{70}) for $r\ll r_{0s}$ and
by formula~(\ref{71}) for $r \gg r_{0s}$. Stream lines
concentrate toward the equatorial plane of a black hole.

To find second order correction $x_{2s}(\theta)$ to the shape of the sonic
surface one should expand the condition $D(x_s)=0$ up to the second order
in~$\epsilon$ similarly to that was done when determining first order
correction~$x_{1s}$~(\ref{61}). However, because the second order
solution in the transsonic region is actually the same as the solution for
$a\ll {\cal M}$ and $r_s \gg {\cal M}$ obtained in BP, it is much
easier to notice that the shape of the sonic surface up to the second
order in~$\epsilon$ coincides with that found in BP for the case $a\ll
{\cal M}$ and ${\cal M}/r_s \ll 1$. So, rewriting formula~(74) from BP and
adding term proportional to~$\epsilon^2$ from expression
for~$x_{sw}$~(\ref{58}), we obtain in our notations
$$
x_{2s}=\epsilon^2\frac{a^2}{{\cal
M}^2}\frac{2(\Gamma-1)}{(5-3\Gamma)^2}
\left[(6\Gamma-10)\cos^2\theta+(k_2(\Gamma)-1)(\Gamma+1)(3\cos^2\theta-1)
\right]+\epsilon^2\frac{27}{16}\frac{(3\Gamma^2-14\Gamma+3)(\Gamma-1)}{
(5-3\Gamma)^3}\mbox{,}
$$
where $\displaystyle k_2(\Gamma)=x_{0s}^3\frac{1}{{\hat a}^2}\left.
\frac{dg_{22}}{dx}\right|_{x=x_{0s}}$ is computed numerically in BP
(see Table~1 there), $k_2(4/3)=0.861$. The shape of the sonic surface is
also shown on Fig.~4 for $\Gamma=4/3$ and $a={\cal M}$.

\section{Conclusions}

Using stream (Grad--Shafranov) equation~(\ref{asc14}) we have considered
the problem of hydrodynamic accretion onto a resting Kerr black hole. The
fact that in astrophysical situations, when this type of accretion could be
realized (accretion of interstellar medium onto an isolated black hole),
the temperature of the surrounding gas is nonrelativistic introduces small
parameter in the problem and allows to obtain analytical solution of the
stream equation~(\ref{asc14}) in the case of an arbitrary value of the
rotational parameter~$a$, extending thus the results of the work BP onto
rapid rotation of the black hole.

Form of the sonic surface and corrections to the particle
density are determined. However, because the sonic surface is located far
from the horizon of a black hole these corrections are very small and flow
picture does not differ drastically from the case of the accretion onto a
Schwarzschild black hole save the dragging into rotation around Kerr black
hole with Lense--Thirring angular velocity. Therefore, the related
astrophysics (emergent radiation) does not change significantly from
previous consideration by Shapiro, 1974. Presence of a magnetic
field or small angular momentum of the gas at infinity relative to the
black hole have much more influence on the accretion pattern and emergent
radiation. So, the results of the present work lie primarily in the scope
of mathematical physics and show that shock--free smooth solution exists
in the problem of the steady state hydrodynamic accretion onto a rotating
black hole with arbitrary $a\leq{\cal M}$.

\begin{center}
{\Large \bf Acknowledgments}
\end{center}

The author is grateful to V.S.~Beskin for fruitful discussions and
providing the results of the work BP before publication. This research is
partially supported by International Science Foundation, grant number
N9X000, and Russian National Scientific Programme "Astronomy", project
number 2--217.

\newpage

\begin{center}
{\Large References}
\end{center}
\parindent0pt

Begelman~M.C., Blandford~R.D., and Rees~M.J., 1984, Rev. Mod. Phys.,
{\bf 56}, 255

Beskin~V.S., Pariev~V.I., 1993, Physics Uspekhi, {\bf 36}, 529

Beskin~V.S., Pidoprigora~Yu.M., 1995, Zhurn. Exp. Teor. Fiz. {\bf 107},
1025~(BP)

Bondi~H., 1952, Mon.~Not. Roy. Astron. Soc., {\bf 112}, 195

Camenzind~M., 1990, Rev.~Mod.~Astron., {\bf 3}, 234

Lipunov~V.M., 1992, Astrophysics of Neutron Stars. Springer--Verlag,
Berlin

Novikov~I.D., Frolov~V.P., 1986, The physics of Black Holes. Nauka Press,
Moscow

Petrich~L.I., Shapiro~S.L., and Teukolsky~S.A., 1988, Phys. Rev. Lett.,
{\bf 60}, 1781

Shapiro~S.L., 1974, Astroph.~J., {\bf 189}, 343

Shapiro~S.L., Teukolsky~S.A., 1983, Black Holes, White Dwarfs,
and Neutron Stars. Wiley--Interscience Publication, New York

Thorne~K.S., Price~R.H., Macdonald~D.A., 1986, Black Holes: The Membrane
Paradigm. Yale University Press, New Haven and London

Zeldovich~Ya.B., Novikov~I.D., 1967, Relativistic Astrophysics.
Nauka Press, Moscow

\vspace{2cm}
\begin{center}
{\Large Figure Captions}
\end{center}
\parindent0pt

{\bf Figure 1.} Functions $y_0(x)$ corresponding to accretion and ejection
solutions of equation~(\ref{36}) for $\Gamma=4/3$.

{\bf Figure 2.} Dependencies of $y_1$ on $x$ for different values
of~$\Gamma$.

{\bf Figure 3.} Function~$G(x)$ for different values of~$\Gamma$.

{\bf Figure 4.} Stream lines and sonic surface for $\epsilon=0.3$,
$a={\cal M}$ and $\Gamma=4/3$.

\end{document}